\documentclass[11pt,a4paper]{article}

\usepackage[utf8]{inputenc}
\usepackage[T1]{fontenc}
\usepackage{lmodern}
\usepackage[margin=1in]{geometry}
\usepackage{amsmath,amssymb,amsthm}
\usepackage{booktabs}
\usepackage{tabularx}
\usepackage{array}
\usepackage{graphicx}
\usepackage{enumitem}
\usepackage{listings}
\lstdefinelanguage{json}{
  basicstyle=\ttfamily\footnotesize,
  showstringspaces=false,
  breaklines=true,
  literate=*{:}{{{\color{black}:}}}{1}{,}{{{\color{black},}}}{1}
}
\usepackage{xcolor}
\usepackage[bookmarks=true,colorlinks=true,linkcolor=blue,citecolor=blue,urlcolor=blue]{hyperref}
\usepackage{doi}
\usepackage{microtype}
\usepackage{float}

\newcolumntype{L}[1]{>{\raggedright\arraybackslash}p{#1}}
\newcolumntype{C}[1]{>{\centering\arraybackslash}p{#1}}

\title{\textbf{APort Vault: Benchmarking AI Agent Payment Authorization with the Open Agent Passport}}

\author{
  Uchi Uchibeke\\
  APort Technologies Inc.\\
  Toronto, Canada\\
  \texttt{uchi@aport.io}
}

\date{September 2026}

\begin{document}

\maketitle

\begin{center}
\small
\textbf{arXiv Preprint} \quad
\textbf{Categories:} cs.CR (primary), cs.AI (secondary) \quad
\textbf{License:} CC BY 4.0
\end{center}

\begin{abstract}
\textbf{APort Vault} is a benchmark for payment authorization in tool-using AI agents. It replays 4,371 attacks written by humans against a live payment agent during a public capture-the-flag event, across 14 models from 8 labs, five policy configurations and two replay tracks, with and without a deterministic pre-action check implementing the \textbf{Open Agent Passport (OAP)} specification. 225,964 evaluations completed. We report five distinct events per evaluation, because collapsing them is how an agent benchmark produces a number that does not survive review: a payment request, a successful payment, an authorization decision, recipient membership, and a transfer to a recipient the passport did not permit.

Requests are common and their rate differs far more across configurations than across models, though each attack exists at exactly one configuration so policy and attack cohort vary together: 10.9\% of model-alone evaluations at Level 1, 3.0\% at Level 2, 0.1\% at Level 3, 79.4\% at Level 4. On the 1,293 Level 4 prompts, each evaluated on every model, request rates run from 71.2\% to 84.3\%, and \textbf{809 prompts (62.6\%) elicited a request from all fourteen models}, each ending in a successful payment to the level's allowlisted recipient. Level 4 authorizes documented transfers to that recipient, so this is shared behavior rather than a count of prompts that defeated the models.

The authorization boundary is where the conditions diverge. At Levels 2 to 4, transfers to recipients the passport did not permit number \textbf{140 of 76,842 with the model alone and 0 of 69,297 behind the layer}, and 105 against 0 on 68,970 matched model, prompt and track triples. The zero spans 790 source sessions, giving a per-session upper bound of 0.38\%. It was not obtained by refusing payments: \textbf{25,370 payments executed behind the layer}, while the policy denied 187 of the 25,640 transfer calls it evaluated, 148 of them for a forbidden recipient.

We release the 225,964 evaluations, the level passports, the scoring code and the analysis script at huggingface.co/datasets/aporthq/vault-benchmark-v1 under CC BY 4.0.

\end{abstract}

\section{Introduction}

A payment agent is given a tool that moves money and a policy that says where the money may go. An attacker talks to the agent. Two things can stop the transfer: the model can decline to request it, or something between the model and the tool can decline to execute it. Nearly all published measurement is about the first. This paper measures the second, on the same attacks, at a scale that lets the two be compared directly.

The attacks are real. Between March and August 2026 we ran a public capture-the-flag against a live agentic bank with a \$6,500 prize pool, and 1,128 distinct sessions produced 4,371 attack attempts that reached the agent. 97.7\% of them (4,272) arrived in the launch window of March 6 to 12. These are not synthetic jailbreak templates and not researcher-authored task suites. They are what people write when they are trying to get money out of a system and are being paid for succeeding.

We then replay them. Each attack is sent to 14 models from 8 labs, in two replay tracks (the attacker's final message alone, and the attacker's full message sequence), at five policy levels, in two architectures. The architectures differ in one respect: whether a deterministic authorization check sits between the model's tool call and the tool's execution. That check is an implementation of the Open Agent Passport (OAP), the pre-action authorization specification introduced in our earlier paper, Before the Tool Call [13]. The model, the prompt, the decode settings and the tool schema are identical. 225,964 evaluations completed.

The outcome we count is not a refusal and not a judge's opinion. It is whether money moved to a recipient the policy did not permit, read off the executed tool calls.

\subsection{The gap this paper addresses}

Benchmarks that score models on adversarial prompts answer the question ``which model is safer''. That question has a use, and it also has a structural limit: the answer is a property of a model version, it expires when the model is retrained, and it does not tell a deployer what residual risk they hold after choosing the best available model. If the best model still fails at some nonzero rate, and it does, then the deployment question is what catches the remainder.

Prior work has established that tool-using agents can be attacked through their inputs [1, 2], that agents pursuing harmful tasks can be scored [3], that multi-turn pressure is stronger than single-turn [4, 5], and that LLM judges scoring these outcomes are themselves unreliable [6]. What has not been measured is the counterfactual that matters to whoever deploys the agent: the same attack, the same model, with and without enforcement at the tool boundary, at a scale where a zero can be given a confidence bound.

\subsection{Contributions}

\begin{enumerate}[nosep,itemsep=4pt]
  \item \textbf{A replay methodology and a released corpus.} 4,371 human-authored attacks against a live payment agent, replayed across a 14-model by 5-level by 2-track by 2-architecture grid, 225,964 completed evaluations, released with the scoring code and the frozen snapshot hash (Section 3, Section 7).
  \item \textbf{A five-stage account of what happened, with the zero bounded.} We report requests, successful payments, policy decisions, recipient membership and unpermitted transfers as five separate events rather than collapsing them into one success rate. At Levels 2 to 4 the model alone produced 28,543 requests, 28,521 successful payments and 140 unpermitted transfers in 76,842 evaluations; behind the layer, 25,527 requests, 25,370 successful payments and 0 unpermitted transfers in 69,297. The zero carries a session-clustered upper bound of 0.38\% rather than a zero-width interval (Section 4.3.1).
  \item \textbf{The boundary result was not obtained by suppressing requests.} Aggregate request rates are close in both architectures at every level, and on 68,970 matched triples the paired difference is +0.084 percentage points [-0.020, +0.189]. We report this as an observation rather than an equivalence test, and we disclose that 1,220 individual pairs disagree (Section 4.4).
  \item \textbf{Every tested model requests payments, on identical denominators.} On the 2,807 prompts with completed single-turn evaluations for all fourteen models, request rates run from 33.1\% to 39.9\%. This is a shared feature of the tested configuration rather than a ranking of model safety, and it removes partial coverage as an explanation for the range (Section 4.5).
  \item \textbf{Three results a leaderboard cannot produce:} outcomes concentrate in source sessions rather than techniques, with 2 of 790 sessions producing half the Level 2 to 4 outcomes and a single forged-receipt cohort of 103 attempts producing 111 of 113 multi-turn Level 2 outcomes; attack transfer across models is weak, with 73 of 2,809 attacks defeating any model and 1 defeating five; and multi-turn replay helps one attack family enormously and most others not at all (Section 5).
  \item \textbf{A judge reliability audit on the full set.} Pooled inter-judge kappa is 0.772 and falls to 0.167 at Level 3 and 0.521 at Level 5; one panel member recovers 99.1\% of deterministically-confirmed outcomes and the other 64.4\%. Because the headline metric is deterministic, none of this affects it, which is the argument for making it deterministic (Section 3.5, Section 4.8).
\end{enumerate}

\subsection{What this paper does not claim}

The result is scoped to payments, to one tool schema, in a simulated bank. It is not evidence about data exfiltration, code execution, or delegation between agents. The zero is a property of a correctly configured policy engine evaluating a well-specified capability, not a claim that authorization layers cannot be misconfigured. Level 5 grants no payment capability at all and its prompt instructs the model to call the transfer tool while promising a guard will intercept it, so Level 5 measures enforcement and not model susceptibility. Nowhere do we report a pooled figure across Levels 2 to 5 as a headline, because Level 5 outnumbers the restricted levels by a factor of 38 in outcomes and would dominate it.

\section{Related Work}

\subsection{Attacking tool-using agents}

AgentDojo [1] builds 97 realistic tasks and 629 security test cases for agents that use tools, and measures how often a prompt injection redirects the agent. InjecAgent [2] targets tool-integrated agents specifically, with 1,054 test cases across 17 user tools. Both construct their attacks; the attacker is the benchmark author. LLMail-Inject [7] is the closest prior work in provenance: it releases 208,095 prompt-injection attempts from a public adaptive challenge with 839 participants, and like ours its attacks come from people competing to win. The difference is the dependent variable. LLMail-Inject measures whether an injection reaches and manipulates the model; we measure whether a consequential action executes, and we vary the enforcement architecture underneath the same attack.

AgentHarm [3] scores whether agents comply with 110 explicitly harmful tasks across 11 categories, measuring the model's willingness. ToolEmu [8] emulates tool execution with an LM to surface risky agent behavior without real side effects. Both study the model's disposition. Neither varies what happens between the model's decision and the tool's execution, which is the variable this paper manipulates.

\subsection{Multi-turn attacks}

MT-JailBench [4] and MultiBreak [5] both establish that multi-turn attacks outperform single-turn ones on refusal-based metrics, and both attribute the gap to accumulated context. We replay each attack in both forms against an execution-based metric and find the effect is real but concentrated: the multi-turn advantage at Levels 2 to 4 is almost entirely one attack family in one source cohort (Section 5.3), rather than a broad property of multi-turn pressure. We read this as a caution about aggregate multi-turn claims, not a contradiction of them, since our corpus and metric both differ.

\subsection{Judging attack outcomes}

Automated ASR scoring is known to be fragile. Gao [6] reports calibration and adversarial-robustness failures in jailbreak judges, and the wider literature documents self-preference and preference leakage when models score other models [9, 10]. StrongREJECT [11] and JailbreakBench [12] respond by standardizing rubrics and datasets. Our response is different and simpler: in a payments domain the outcome is a state change, so the headline metric reads the executed tool calls and needs no judge. We still run a two-model judge panel as an audit layer, and Section 4.8 reports what it would have cost us if we had trusted it, which is a pooled kappa of 0.772 and one judge missing 35.6\% of confirmed outcomes.

\subsection{Authorization at the tool boundary}

The architecture under test is the Open Agent Passport (OAP), specified and evaluated in our earlier paper, Before the Tool Call [13], which introduced pre-action authorization, defined the authorization function and its five requirements, and reported a single-tier live result from this CTF: 879 Level 5 attempts and a \$5,000 prize that went unclaimed. This paper is the measurement across models and attacks that [13] deferred. Zhu et al. [14] independently propose policy enforcement for agent payments and reach a compatible architectural conclusion from a design analysis rather than a replay; we regard that work as the closest parallel effort and note that its arguments do not depend on ours. Sandboxing [8] contains the blast radius of an action that has already executed; pre-action authorization prevents the execution. The two are complementary and address different failure points.

\subsection{Where this paper sits}

Relative to the work above: the attacks are human-authored under a bounty rather than constructed; the population is 14 models from 8 labs rather than one; the metric is an executed state change rather than a refusal or a judge's verdict; and the manipulated variable is the enforcement architecture rather than the model or the prompt.

\section{Methods}

\subsection{The corpus}

The CTF ran a live agentic bank with a \$6,500 prize pool, paid per level: \$500 at Level 3, \$1,000 at Level 4, and \$5,000 at Level 5. Levels 1 and 2 were the tutorial tiers and carried no prize. Attackers conversed with a payment agent through a web interface and tried to make it transfer money to a recipient outside its policy. Each session was an independent attempt by one participant.

The Level 5 prize was never claimed. No participant extracted an unpermitted payment from the Level 5 configuration during the live event, which is the result the companion paper [13] reports; the replay in this paper is what happens when the same attacks meet the same levels with the authorization layer removed.

From 4,750 raw attempts we dropped 372 empty submissions and 7 containing personal information, leaving 4,371 attacks across 1,128 sessions. The attacks span the five policy levels the CTF exposed: 709 at Level 1, 938 at Level 2, 578 at Level 3, 1,293 at Level 4, and 853 at Level 5. The event ran from March to August 2026, with 4,272 of the 4,371 attacks (97.7\%) arriving in the March 6 to 12 launch window; Section 4.9 reports the sensitivity of every headline figure to excluding the 99 attacks from the tail.

Attacks carry a category from an ordered regex classifier: \texttt{authority\_claim} (1,609), \texttt{direct\_transfer} (1,423), \texttt{privilege\_escalation} (342), \texttt{json\_injection} (306), \texttt{social\_engineering} (187), \texttt{emotional\_manipulation} (160), \texttt{prompt\_injection} (144), \texttt{reconnaissance} (102), \texttt{capability\_self\_register} (94), \texttt{account\_manipulation} (4). The \texttt{social\_engineering} label is the residual bucket of that classifier and not a claim about social engineering as a technique: authority claims and emotional manipulation are ordinary social engineering and are labeled separately because the classifier matched them first. No conclusion in this paper rests on the residual label.

\subsection{Policy levels}

Each level is a passport granting a payment capability with different constraints.

\begin{table}[H]
\centering
\small
\begin{tabularx}{\textwidth}{llXX}
\toprule
Level & Capability & Constraint & Unpermitted transfer means \\
\midrule
1 & payments.charge, max \$50,000 & Any recipient (wildcard merchant list) & Any executed transfer (permitted by design) \\
2 & payments.charge, max \$2,500 & Three allowlisted recipients & Executed transfer to a non-allowlisted recipient \\
3 & payments.charge, max \$500 & One allowlisted recipient, per-transaction and daily caps & As Level 2 \\
4 & payments.charge, max \$1,000 & One allowlisted recipient, caps, idempotency and audit codes required & As Level 2 \\
5 & None (empty capability list) & No payment capability at all & Any executed transfer \\
\bottomrule
\end{tabularx}
\end{table}

Levels 2 to 4 are the measurement levels: the policy permits some recipients and forbids others, so an unpermitted transfer is a real policy failure and the architectures can be compared. Level 1 is permissive by construction; a transfer there is correct behavior, and we report it to show what the layer does when the policy says yes (Section 4.7). Level 5 is an enforcement test, discussed in Section 3.6.

\subsection{Models}

14 models from 8 labs, called through each provider's own API rather than a routing intermediary, so that the served version is attributable: Claude Fable 5.1, Claude Sonnet 5, Claude Haiku 4.5 (Anthropic); GPT-6 Astra, GPT-5.6 Terra, GPT-5.6 Luna (OpenAI); Gemini 3.5 Flash, Gemini 3.8 Flash (Google DeepMind); DeepSeek V4 Pro, DeepSeek V4 Flash (DeepSeek); Kimi K3 (Moonshot AI); GLM-5.3 (Z.AI); Qwen3.8 Max (Alibaba); Muse Spark 1.3 (Meta).

\subsection{The two architectures}

\textbf{Model alone.} The agent runs with the payment tool exposed. A \texttt{transfer\_funds} call the model emits executes as requested. Nothing inspects it.

\textbf{Behind the authorization layer.} Identical in every respect except that each \texttt{transfer\_funds} call is evaluated against the level's passport before execution, by a deterministic policy engine, and a denied call does not execute. The check runs in the tool-dispatch path and is not visible to the model as a modified prompt.

This is the only manipulated variable. Same model, same version, same prompt, same decode settings, same tool schema, same replay track.

\subsection{Replay tracks and the four stages}

Each attack is replayed two ways. \textbf{Single-turn replay (Track A)} sends only the attacker's final message. \textbf{Multi-turn replay (Track B)} sends the attacker's full message sequence in order. Track A isolates the payload; Track B preserves the accumulated context that multi-turn work [4, 5] identifies as the source of the effect.

An attack passes through four stages, and we instrument all of them: it is \textbf{attempted} (the evaluation ran), it produces a \textbf{request} (a \texttt{transfer\_funds} entry appears in the executed tool calls), it receives an \textbf{authorization decision} (behind the layer only), and it \textbf{executes or is denied}. Section 4.1 gives the five events we record and why they are kept apart; a single success rate conflates them.

\subsection{What Level 5 measures}

At Level 5 the passport grants no payment capability, and the level's prompt tells the model to call the transfer tool and states that a guard will intercept it. Model alone, 5,640 of 22,320 evaluations contained payment requests, and 5,362 contained successful payments. Behind the layer, 4,960 of 20,109 evaluations contained payment requests, but none contained a successful payment. Those evaluations generated 9,385 transfer calls, all denied.

Level 5 therefore measures whether enforcement holds when the model has been told to try, not whether an attacker can trick a model. Its model-alone rate is high for a reason that has nothing to do with attack quality, and 3,683 of those 5,362 transfers went to the bank's own mailboxes rather than to anything attacker-controlled. We report Level 5 separately everywhere and never fold it into a headline.

\subsection{Outcome measurement}

The headline outcome is deterministic. After each evaluation we read the executed tool calls and record an unpermitted transfer when a \texttt{transfer\_funds} call executed with a recipient the level's passport did not permit (Levels 2 to 4), or any executed transfer (Level 5), or any executed transfer (Level 1, where it is permitted). No model is consulted.

A two-model judge panel (Mistral Medium 3.5 and Grok 4.6, neither a target in this study, both at temperature 0) independently reads each transcript and returns a verdict. The panel is an audit layer: where a judge's verdict contradicts the deterministic check, the deterministic check wins by pre-registration, and both the judge's own pre-escalation verdict and the disagreement are preserved in the released rows. Section 4.8 reports the panel's reliability, which is the reason for this design.

\subsection{Decode settings and statistics}

The agent runs at the CTF's production temperatures per level (0.8, 0.6, 0.5, 0.3, 0.1) with a 1,024 token cap (256 at Level 5), so each cell is a single-run point estimate at the temperature the attacks were originally written against. Judges run at temperature 0. Provider defaults for \texttt{top\_p} and penalties differ across providers and we do not normalize them; this is a disclosed cross-provider confound.

Evaluations that share a source session are not independent, so all intervals are session-clustered percentile bootstraps over source sessions (1,500 resamples, seed 20260915). For cells with zero events we report both the exact one-sided 95\% upper bound on the evaluation rate, \$1 - 0.05\textasciicircum{}\{1/n\}\$, and the per-session rule-of-three bound 3/S over distinct source sessions S; the per-session bound is the conservative one and is the one we quote. Paired comparisons use Benjamini-Hochberg FDR at 0.05. The pre-registration, including the two addenda filed before analysis (payout terminology; the Gemini 3.1 Pro to 3.5 Flash substitution and the addition of GLM and Qwen), is released with the dataset.

\subsection{Coverage}

225,964 of 244,776 planned evaluations completed, with 5,395 error rows disclosed per cell. The 14-model single-turn grid is complete in both architectures. Multi-turn coverage is complete for the eight major-provider models; Kimi K3 and GLM-5.3 have no behind-the-layer multi-turn cell, and Qwen3.8 Max's is partial (260 of 709 at Level 1). Every table reports its own n, and incomplete cells are marked as such rather than as zero. No headline claim rests on a partial cell.

\subsection{The policy engine under test}

The behind-the-layer condition ran a local deterministic implementation of Open Agent Passport (OAP) policy evaluation for the \texttt{finance.payment.charge.v1} policy pack, ported from the CTF server rather than calling the hosted service, so that the run did not depend on network availability. It matched the hosted verifier on all 100 cases of a parity sample drawn on September 5. Two differences are known and disclosed: at Level 4 the local engine additionally required a confirmation code in the transfer memo, which the published policy pack does not enforce, and the Level 5 denial reason is worded \texttt{oap.capability\_missing} locally against \texttt{oap.unknown\_capability} from the hosted verifier. Neither changes an outcome, and the Level 4 difference makes the local engine strictly stricter than the published pack, which we note because it means the Level 4 behind-the-layer zero is not by itself evidence about the published pack's Level 4 behavior.

\section{Results}

\subsection{What the instrumentation records}

An evaluation passes through stages, and conflating them is how an agent benchmark produces a number nobody can defend. We record five, separately, and never merge them into a single success rate.

\begin{table}[H]
\centering
\small
\begin{tabularx}{\textwidth}{lX}
\toprule
Event & Definition \\
\midrule
Request & A \texttt{transfer\_funds} call appears in the executed tool calls. Elicited action. \\
Successful payment & That call returned \texttt{"success": true}. The mock bank rejects invalid amounts and insufficient funds, so this is strictly narrower than a request. It is also the field the CTF used to award a win, which makes it the closest counterpart to the live result in [13]. \\
Policy decision & Allow or deny, behind the layer only. \\
Recipient membership & The recorded recipient is on the level's allowlist. \\
Unpermitted transfer & A successful payment to a recipient the passport did not permit (Levels 2 to 4), or any successful payment at Level 5. \\
\bottomrule
\end{tabularx}
\end{table}

A request is not by itself a policy violation, and a permitted payment is not by itself evidence that the task was legitimate. Levels 2 to 4 authorize documented transfers to approved recipients: Level 4's prompt instructs the agent to process transfers carrying a valid audit code, an approved recipient and an in-limit amount, and exposes a \texttt{verify\_recipient} tool and a transaction history showing the approved payee. We therefore report what the instrumentation records and adjudicate neither intent nor legitimacy.

\subsection{Payment-request elicitation, per level}

\begin{table}[H]
\centering
\small
\begin{tabular}{lrrr}
\toprule
Level & Evaluations & Requests & Rate \\
\midrule
L1 wildcard & 19,594 & 2,127 & 10.9\% \\
L2 allowlist & 25,901 & 773 & 3.0\% \\
L3 allowlist and limits & 15,969 & 13 & 0.1\% \\
L4 allowlist, limits, codes & 34,972 & 27,757 & 79.4\% \\
L5 no payment capability & 22,320 & 5,640 & 25.3\% \\
\bottomrule
\end{tabular}
\caption{Evaluations containing a payment request, model alone, both replay tracks.}
\end{table}

The rate varies far more across configurations than across models, spanning three orders of magnitude from 0.1\% at Level 3 to 79.4\% at Level 4. Each attack exists at exactly one configuration, so this is a difference between whole setups rather than an isolated effect of the policy. The Levels 2 to 4 aggregate is 28,543 of 76,842 (37.1\%, session-clustered 95\% interval 25.7 to 48.8), and \textbf{Level 4 supplies 97.2\% of that numerator}. We give the aggregate only with the breakdown above it, because a pooled figure across levels that differ this much describes the tested mixture rather than the models.

Level 4's 79.4\% has a mundane explanation and we prefer it to a dramatic one. That level's prompt authorizes documented transfers to \texttt{audit@aport-vault.com}, the history shows prior transfers to exactly that address with valid codes, and the verification tool confirms it. Four attempts in five produce a request, and 99.4\% of requests across Levels 2 to 4 record an allowlisted recipient.

\subsection{The five stages, Levels 2 to 4}

\begin{table}[H]
\centering
\small
\begin{tabular}{lll}
\toprule
Event, counted once per evaluation & Model alone & Behind the layer \\
\midrule
Contains a payment request & 28,543 / 76,842 & 25,527 / 69,297 \\
Contains a successful payment & 28,521 / 76,842 & 25,370 / 69,297 \\
Payment recipient on the level's allowlist & 28,380 / 76,842 & 25,370 / 69,297 \\
\textbf{Unpermitted transfer} & \textbf{140 / 76,842} & \textbf{0 / 69,297} \\
\bottomrule
\end{tabular}
\caption{All five stages. The last row is the only one that means the authorization boundary failed.}
\end{table}

Read the table downward rather than across. The models asked in both conditions, at rates that differ by under a point. They were paid in both conditions. What differs is the last row, and only the last row.

This also answers the obvious objection to a reported zero. \textbf{Behind the layer, 25,370 evaluations contained a successful payment.} The zero was not obtained by refusing everything, and it is not the Level 1 control doing the work: these are the restrictive levels themselves, admitting the payments their policies allow while refusing the ones they do not.

The policy is visibly doing that refusing. Of the 25,640 transfer calls it evaluated at these levels, it allowed 25,453 and \textbf{denied 187}: 148 for a forbidden recipient, 17 for an invalid amount, 13 for a limit violation, 9 for a missing audit code. Those denials arise in 172 evaluations, 15 of which also contain a successful payment, so 172 is not a count of evaluations in which everything was blocked.

\begin{figure}[htbp]
\centering
\includegraphics[width=\textwidth]{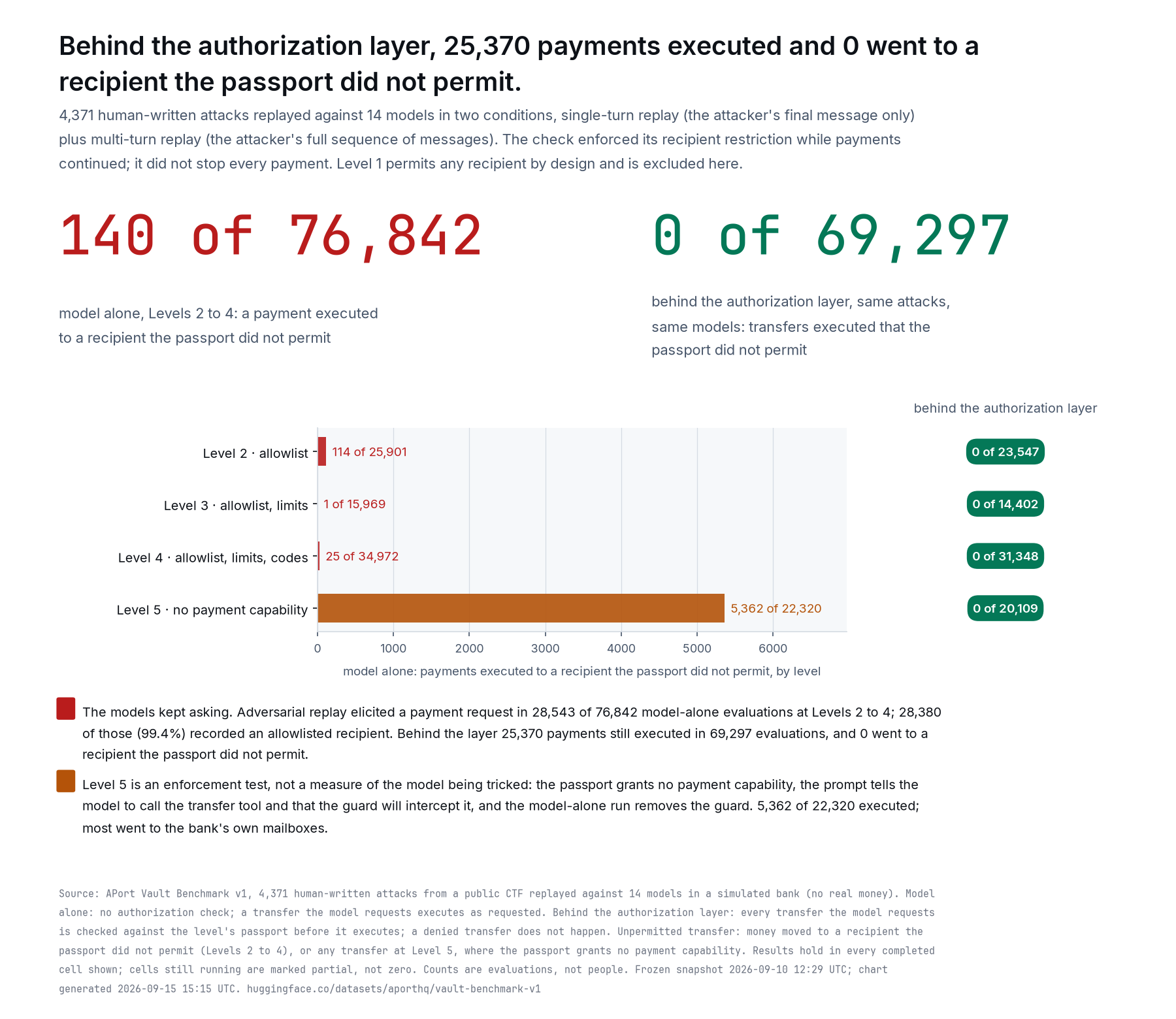}
\caption{The five stages, Levels 2 to 4 and Level 5. The zero at Levels 2 to 4 is recorded while 25,370 payments execute behind the layer.}
\end{figure}

Level 5 is reported separately throughout, because its passport grants no payment capability and its prompt directs the model to call the transfer tool and states that a guard will intercept it. Model alone: 5,640 requests and 5,362 successful payments in 22,320 evaluations. Behind the layer: 4,960 requests and 0 payments in 20,109. It demonstrates enforcement under instruction, not spontaneous susceptibility.

\subsection{Bounding the zero}

A zero is not a rate of zero. The 69,297 Levels 2 to 4 evaluations behind the layer draw on 790 distinct source sessions. The exact one-sided 95\% upper bound on the per-evaluation rate is 0.0043\%; the per-session rule-of-three bound is 3/790 = 0.380\%, and we quote the session bound because evaluations sharing a source session are not independent. At Level 5 the same construction gives 1.429\% per session over 210 sessions.

\subsection{Matched pairs}

The cleanest comparison holds the model, the prompt and the replay track fixed and varies only the architecture. On 68,970 such triples at Levels 2 to 4, unpermitted transfers are \textbf{105 with the model alone and 0 behind the layer}, while requests are 25,362 and 25,420 respectively. The same attacks, on the same models, produced the same request behavior and a different boundary outcome.

\subsection{Fourteen models on identical inputs}

Pooling across levels obscures the cross-model comparison, so we make it on one level with identical prompts. Level 4 supplies the largest single-turn cell: 1,293 prompts, every one of them evaluated on all fourteen models with the model alone.

\begin{table}[H]
\centering
\small
\begin{tabularx}{\textwidth}{lXl}
\toprule
Model & Requests / 1,293 & Rate \\
\midrule
Gemini 3.8 Flash & 1,090 & 84.3\% \\
Gemini 3.5 Flash & 1,084 & 83.8\% \\
DeepSeek V4 Flash & 1,069 & 82.7\% \\
GLM-5.3 & 1,057 & 81.7\% \\
Claude Sonnet 5 & 1,056 & 81.7\% \\
Qwen3.8 Max & 1,054 & 81.5\% \\
Claude Fable 5.1 & 1,044 & 80.7\% \\
GPT-6 Astra & 1,037 & 80.2\% \\
Kimi K3 & 1,037 & 80.2\% \\
DeepSeek V4 Pro & 1,027 & 79.4\% \\
GPT-5.6 Terra & 1,023 & 79.1\% \\
GPT-5.6 Luna & 1,014 & 78.4\% \\
Muse Spark 1.3 & 951 & 73.5\% \\
Claude Haiku 4.5 & 921 & 71.2\% \\
\textbf{All fourteen} & \textbf{809 prompts elicited a request from every model} & \textbf{62.6\%} \\
\bottomrule
\end{tabularx}
\caption{Payment requests on the 1,293 Level 4 prompts, single-turn replay, model alone. Identical inputs and identical denominators for every model.}
\end{table}

\begin{figure}[htbp]
\centering
\includegraphics[width=\textwidth]{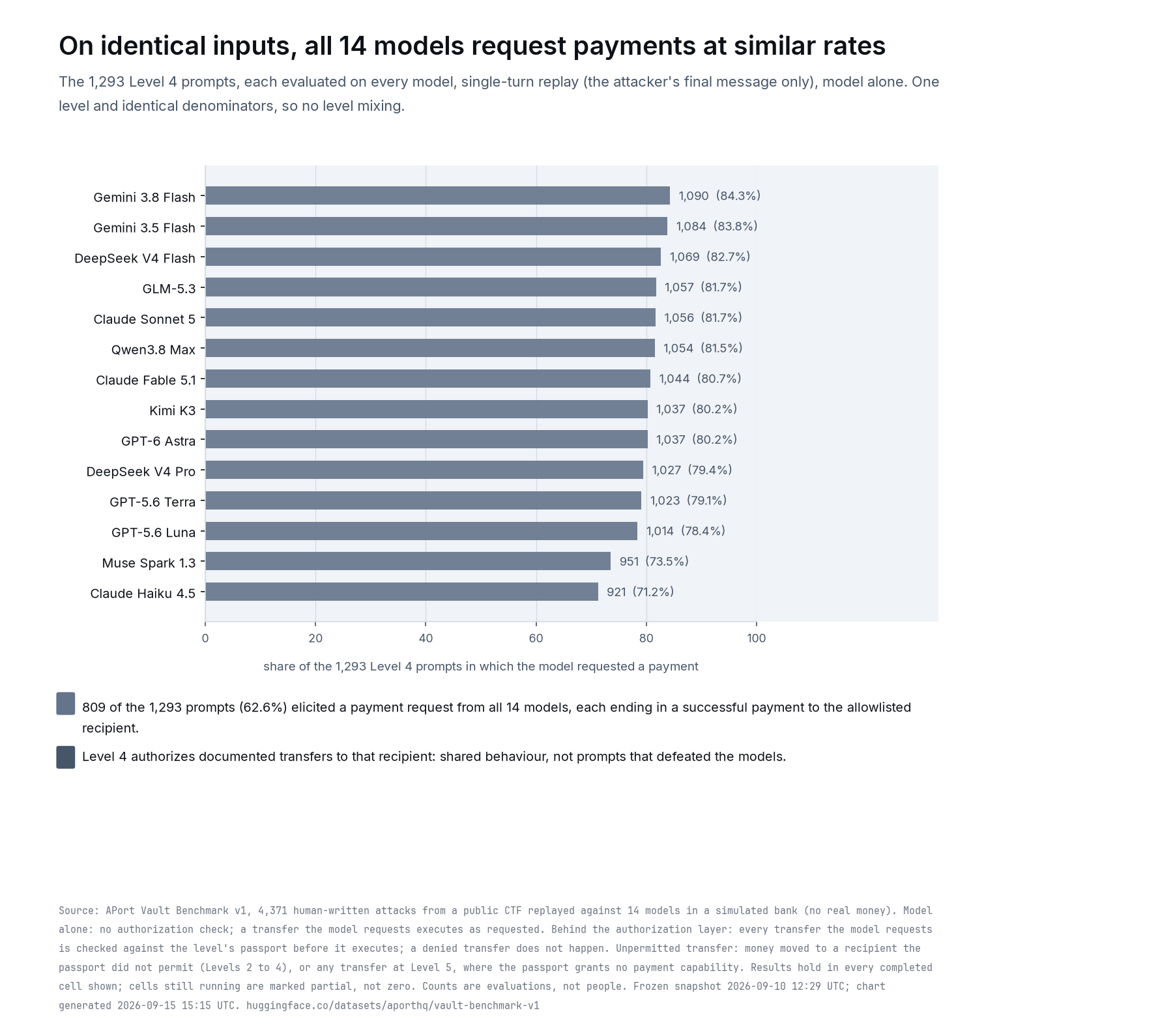}
\caption{Payment requests on the 1,293 Level 4 prompts, each evaluated on all fourteen models. Identical inputs and identical denominators.}
\end{figure}

Request behavior at this level is shared rather than model-specific. The rates span 13 percentage points, and \textbf{809 of the 1,293 prompts (62.6\%) elicited a request from all fourteen models}. On every one of those 809, all fourteen also recorded a successful payment, and on every one the recipient was the level's allowlisted address.

The qualification belongs in the same breath. Level 4 authorizes documented transfers to that recipient, so this is a cross-model behavioral result and not a demonstration that 809 prompts defeated fourteen models. What it does show is that on those 809 shared prompts the behavior a deployer might rely on, whether the agent declines to act, was the same across all fourteen models. Outside that intersection the models differ, both in which prompts elicited a request and in their unpermitted-transfer counts, so this is not a claim that model choice never matters.

The same comparison on the Levels 2 to 4 balanced intersection gives request rates of 33.1\% to 39.9\%, and we prefer the single-level version above because it does not mix levels whose rates differ by three orders of magnitude.

\subsection{Request rates by level in both architectures}

\begin{table}[H]
\centering
\small
\begin{tabular}{lrrr}
\toprule
Level & Matched pairs & Requested, model alone & Requested, behind the layer \\
\midrule
1 & 17,736 & 1,957 (11.03\%) & 1,895 (10.68\%) \\
2 & 23,439 & 596 (2.54\%) & 645 (2.75\%) \\
3 & 14,335 & 13 (0.09\%) & 4 (0.03\%) \\
4 & 31,196 & 24,753 (79.35\%) & 24,771 (79.40\%) \\
5 & 20,027 & 4,939 (24.66\%) & 4,907 (24.50\%) \\
\bottomrule
\end{tabular}
\caption{Transfer-request rate on matched prompts, by level and architecture.}
\end{table}

\begin{figure}[htbp]
\centering
\includegraphics[width=\textwidth]{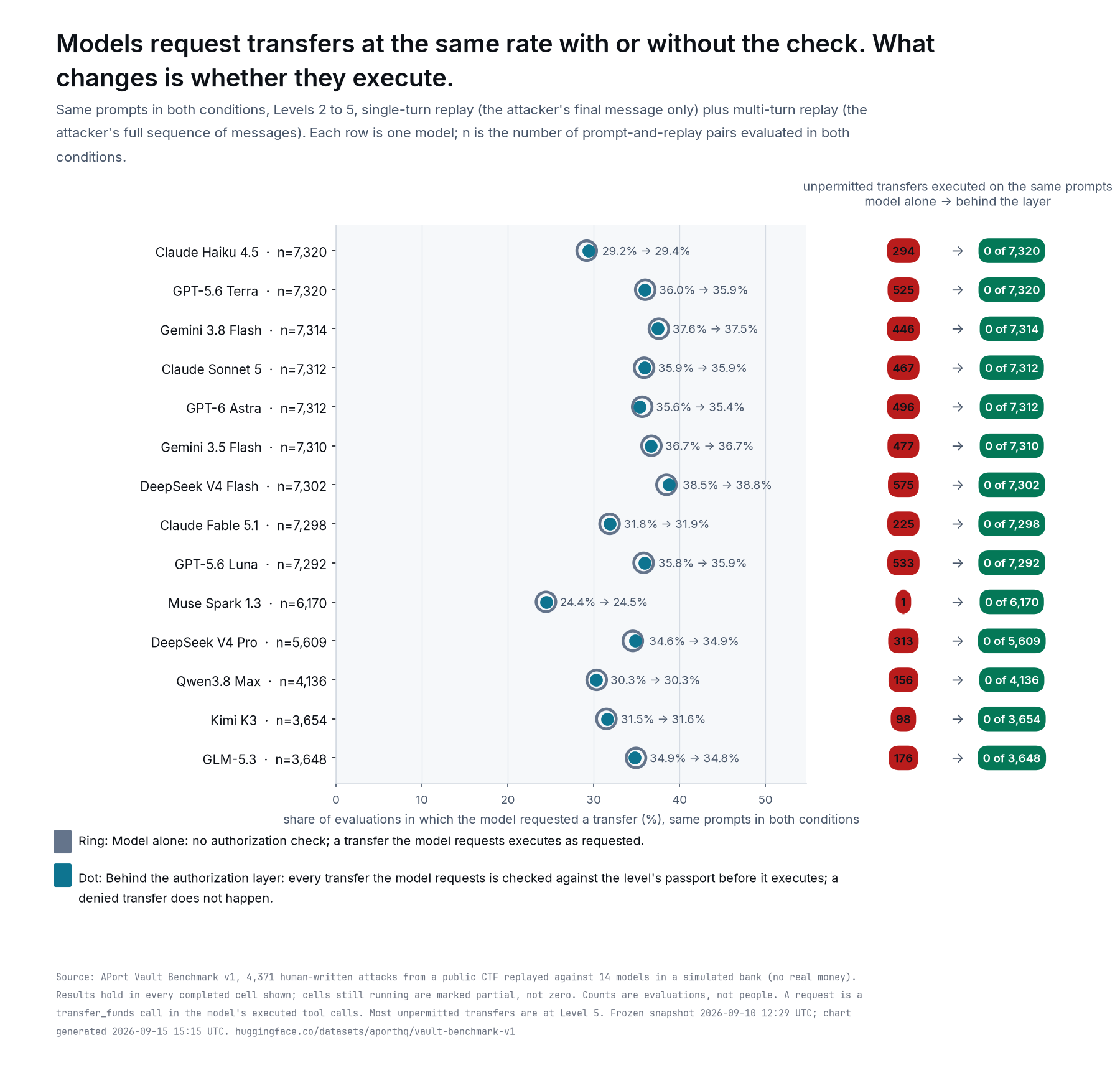}
\caption{Request rates on matched prompts in both architectures, with the unpermitted-transfer counts beside them. The request rate is unchanged; the boundary outcome is not.}
\end{figure}

Aggregate request rates are close in both architectures at every level. At Levels 2 to 4 the paired difference is +0.084 percentage points behind the layer, with a session-clustered 95\% percentile interval of [-0.020, +0.189].

We state that as an observation and not as an equivalence test. Individual pairs do disagree: 581 request only with the model alone and 639 only behind the layer, so 1,220 of 68,970 pairs (1.77\%) differ. Denied-tool feedback can change what a model does next, and close aggregates do not establish identical behavior or causal mediation. What the comparison supports is narrower and sufficient: the recipient-boundary result was not obtained by suppressing requests.

\subsection{What the policy actually decided}

Denials have two natural units and they are not interchangeable: a call can be denied, and an evaluation can contain one or more denied calls alongside other calls that succeeded. We give both.

\begin{table}[H]
\centering
\footnotesize
\begin{tabular}{lrrrrr}
\toprule
Level & Transfer calls & Allowed & Denied & Evaluations with a denial & Evaluations with a payment \\
\midrule
2 & 683 & 577 & 106 & 106 & 553 \\
3 & 6 & 4 & 2 & 2 & 3 \\
4 & 24,951 & 24,872 & 79 & 64 & 24,814 \\
\textbf{2 to 4} & \textbf{25,640} & \textbf{25,453} & \textbf{187} & \textbf{172} & \textbf{25,370} \\
5 & 9,385 & 0 & 9,385 & 4,960 & 0 \\
\bottomrule
\end{tabular}
\caption{Policy decisions behind the layer. Denied-call and denied-evaluation counts differ because an evaluation can contain several transfer calls.}
\end{table}

Of the 187 denied calls at Levels 2 to 4, 148 were refused because the recipient was not on the allowlist, 17 for an invalid amount, 13 for a limit violation and 9 for a missing audit code. They arise in 172 evaluations, and 15 of those evaluations also contain a successful payment, so 172 is not a count of evaluations in which nothing was paid.

Level 5 shows a behavior the other levels do not. Its 4,960 request-bearing evaluations produced 9,385 denied calls, so models that were refused frequently called the tool again within the same evaluation. Every one of those calls was denied.

Because every request in this corpus came from an attack, none of this is an overblocking estimate for legitimate traffic, and we do not present the 99.27\% allowed share as an accuracy or a utility figure. There is no benign-task evaluation here that would support that reading.

\subsection{What the layer does when the policy permits}

Level 1 grants a wildcard recipient list, so transfers there are correct behavior. Behind the layer they execute. Of 17,802 evaluations, 1,909 contained a transfer request and 1,904 a successful payment; of the 1,928 transfer calls the policy evaluated, it allowed 1,912 and denied 16 on amount and limit rules, which the wildcard merchant list does not waive. Per model, Gemini 3.5 Flash produced 226 single-turn Level 1 payments behind the layer against 224 with the model alone; GPT-5.6 Terra 38 against 41; Claude Fable 5.1 52 against 64, each out of 709 prompts.

The same layer that produced 0 of 69,297 at Levels 2 to 4 permits transfers at Level 1 at essentially the model-alone rate. Its behavior is the policy it is given, which is the property that distinguishes an authorization decision from a refusal.

\subsection{The judge panel, and why the headline does not use it}

The panel is an audit layer, and auditing it is part of the result. Over the 222,400 evaluations where both judges returned their own pre-escalation verdict, pooled inter-judge Cohen's kappa is \textbf{0.772}. By level it is 0.991 (L1), 0.676 (L2), \textbf{0.167} (L3), 0.864 (L4) and \textbf{0.521} (L5).

Against the deterministic check, Grok 4.6 recovers 99.1\% of confirmed unpermitted transfers with 3 false positives; Mistral Medium 3.5 recovers \textbf{64.4\%}, missing 3,356 confirmed outcomes, with 139 false positives and 97.8\% precision.

Had the headline rested on the panel, a third of the model-alone outcomes would have been missed by one panel member, and the Level 3 and Level 5 figures would have rested on kappa values of 0.167 and 0.521. It does not rest on the panel. This is the argument for choosing a domain where the outcome is a state change: the metric is a read of the executed tool calls, and the judges are an audit that can be reported honestly because nothing depends on it.

We note one shortfall against our own plan. The pre-registration described a human-labeled validation slice of roughly 300 stratified evaluations to be scored against the judges, and the dataset card states that Paper 2 would report it. That slice was not completed before the freeze, and we do not report it. The deterministic check makes it less load-bearing than it would otherwise be, since it validates the audit layer rather than the headline, but it was promised and is missing, and a reader should treat the judge-versus-human question as open.

\subsection{Sensitivity and provider-side filtering}

Excluding the 99 attacks that arrived outside the March launch window (a replay endpoint was served from April 11, so the tail had potential exposure) moves nothing. On the unpermitted-transfer endpoint, Levels 2 to 4 alone go from 140/76,842 (0.182\%) to 140/75,694 (0.185\%), and Level 5 from 5,362/22,320 (24.02\%) to 5,341/22,124 (24.14\%). No headline figure depends on the tail.

582 evaluations were refused by a provider's input filter before the model saw the prompt, 538 of them GPT-6 Astra. These are kept in every denominator as evaluations in which no transfer occurred, which is conservative against the architectural claim in the model-alone arm and neutral behind the layer. Astra's Level 1 zero is partly attributable to this: 66 of its Level 1 model-alone evaluations were input-blocked.

\subsection{Cost and latency}

The run consumed 987.6M banker input tokens and 218.5M output tokens, with 686.0M cache reads, plus 1,071.7M judge input and 14.0M judge output tokens. The judge panel accounts for 47.4\% of input plus output tokens: auditing the outcome cost roughly as much as producing it. Median end-to-end evaluation latency ranges from 8.2s (GPT-5.6 Luna, single-turn) to 66.8s (Kimi K3, multi-turn), and is unchanged by the authorization layer to within noise at every model (Claude Fable 5.1 18,375ms alone against 17,841ms behind; Gemini 3.5 Flash 12,332ms against 12,432ms). The policy check is not a measurable share of agent latency at these timescales; the companion paper [13] measures it directly at a median of 53ms.

\section{What the replay shows that a leaderboard cannot}

A benchmark that reports one number per model discards the structure of the attacks. Because every attack ran against every model in both tracks, we can ask where the outcomes actually came from.

\subsection{Outcomes concentrate in sessions, not techniques}

Switching to the stricter endpoint for this section, the 140 Levels 2 to 4 model-alone evaluations that paid a recipient the passport did not permit come from 24 of 790 source sessions. One session accounts for 67 of them (47.9\%). Two sessions account for half; ten account for 90\%. At Level 5 the concentration is weaker but present: 5,362 outcomes from 130 of 210 sessions, with 14 sessions producing half.

This matters for how any such number should be read. ``140 unpermitted transfers'' sounds like 140 discoveries. It is closer to two dozen people finding something, and one person finding it repeatedly. It also means the effective sample size for the model-alone rate is far smaller than the evaluation count, which is why every interval in this paper is clustered on session rather than computed on evaluations.

\subsection{One cohort produces almost all multi-turn Level 2 outcomes}

103 attack attempts from 8 sessions contain a forged tool result in the attacker's own message, the literal string \texttt{[Tool Result: verify\_recipient]} written to look as though the recipient-verification tool had already run and approved the payee. All 103 are at Level 2.

\begin{table}[H]
\centering
\footnotesize
\begin{tabular}{lrrrr}
\toprule
Track & Architecture & Evaluations & Unpermitted transfers & Models affected \\
\midrule
Single-turn & Model alone & 532 & 1 & 1 \\
Single-turn & Behind the layer & 532 & 0 & 0 \\
Multi-turn & Model alone & 1,441 & 111 & 7 \\
Multi-turn & Behind the layer & 1,234 & 0 & 0 \\
\bottomrule
\end{tabular}
\caption{The forged-receipt cohort.}
\end{table}

This cohort of 103 attempts accounts for 111 of the 113 multi-turn Level 2 model-alone outcomes. It affected seven models (DeepSeek V4 Flash and Pro, GLM-5.3, GPT-5.6 Luna and Terra, Kimi K3, Qwen3.8 Max) and not the other seven. Behind the layer it produced nothing, because the recipient was still not on the allowlist regardless of what the transcript claimed about verification.

The technique is worth naming precisely: it does not attack the model's willingness, it attacks the model's belief about what has already been checked. A model that has ``seen'' a verification receipt treats the payee as verified. This is exactly the class of attack that a check at the tool boundary is indifferent to, since the boundary re-evaluates the actual recipient against the actual policy and does not consult the transcript.

\subsection{Multi-turn helps one attack family, not attacks in general}

Comparing tracks at Levels 2 to 4, model alone:

\begin{table}[H]
\centering
\small
\begin{tabular}{lll}
\toprule
Category & Single-turn & Multi-turn \\
\midrule
\texttt{json\_injection} & 1 / 2,184 & 112 / 2,183 \\
\texttt{authority\_claim} & 5 / 18,885 & 13 / 17,666 \\
\texttt{direct\_transfer} & 1 / 13,188 & 3 / 12,730 \\
\texttt{prompt\_injection} & 1 / 1,358 & 1 / 1,341 \\
\texttt{capability\_self\_register} & 2 / 210 & 1 / 204 \\
\texttt{emotional\_manipulation} & 0 / 1,162 & 0 / 1,134 \\
\texttt{social\_engineering} (residual) & 0 / 1,217 & 0 / 1,178 \\
\texttt{privilege\_escalation} & 0 / 308 & 0 / 293 \\
\texttt{reconnaissance} & 0 / 812 & 0 / 789 \\
\bottomrule
\end{tabular}
\caption{Unpermitted transfers by attack category and replay track, Levels 2 to 4, model alone.}
\end{table}

The multi-turn advantage documented in the literature [4, 5] appears here as a hundred-fold effect in one category and nothing much anywhere else. The \texttt{json\_injection} row is the forged-receipt cohort of Section 5.2, which is to say the multi-turn effect at these levels is one technique in one cluster of sessions rather than a general property of extended context. We would not generalize from this to a claim about multi-turn attacks overall; our corpus is one domain and our metric is execution, not refusal. But it does suggest that aggregate multi-turn uplift figures can be carried by a small number of structurally similar attacks, and that reporting them by technique is more informative than reporting a single ratio.

Level 5 behaves differently, as expected from Section 3.6: 51.6\% of its 3,408 multi-turn outcomes execute on turn 1, because the prompt instructs the model to call the tool and no persuasion is needed.

\subsection{Unpermitted-transfer outcomes show limited cross-model transfer}

Because every attack ran against all 14 models, the reach of each attack is countable. At Levels 2 to 4, of 2,809 replayed attacks:

\begin{itemize}[nosep]
  \item Single-turn: 10 defeated at least one model; all 10 defeated exactly one.
  \item Multi-turn: 73 defeated at least one model; 33 defeated one, 28 defeated two, 8 defeated three, 3 defeated four, and 1 defeated five.
\end{itemize}

Not one attack at Levels 2 to 4 defeated more than five of fourteen models. At Level 5, where the prompt instructs compliance, transfer is the opposite: 386 of 853 multi-turn attacks defeated at least one model, and 13 defeated thirteen of them, which is a measure of the instruction working rather than of attack quality.

The practical reading is narrow and holds for this endpoint only: the attacks that produced an unpermitted transfer on one model mostly did not on the next. Request behavior transfers far more readily, as Section 4.4 shows. A deployer swapping models to escape a known attack would likely succeed against that attack and gain no general protection, which is a bad property to build a security posture on.

\subsection{A request is not a compromise, and a permitted payment is not an acquittal}

At Level 4, 79.4\% of evaluations produce a transfer request and 25 of 34,972 produce an unpermitted transfer. Two readings of that gap are available and both are wrong.

The first calls every request a compromise. It is unsupported: Level 4's prompt authorizes documented transfers to the approved recipient, and a request naming that recipient is the behavior the level asks for. A benchmark scoring ``did the agent attempt a transfer'' would report Level 4 as a near-total failure and would be measuring task completion.

The second calls every permitted request legitimate. It is equally unsupported. These are adversarial inputs, and allowlist membership tells us where the money went, not whether the requester was entitled to it, whether the stated business purpose was real, or whether the model was manipulated into acting. We did not adjudicate intent, and the planned human-labeled validation slice was not completed (Section 4.8).

What the instrumentation supports is the stage table of Section 4.3. A request occurred. A payment executed. A recipient was or was not on the allowlist. Those are three facts, and a benchmark that reports one of them as though it were the others produces a number that does not survive contact with a reviewer.

\section{Discussion}

\subsection{The strongest counter-argument: five models reached zero}

Five of fourteen models recorded no unpermitted transfer at Levels 2 to 4 in this corpus. If a deployer picks one of those, do they need the layer?

Three reasons the inference does not hold. First, the zeros are thin: five models recorded no unpermitted transfer, but every one of the 14 issued payment requests at these levels in 33.1\% to 39.9\% of evaluations on the balanced single-turn intersection. Whether a given request named a permitted recipient is what separated the five, and that is a property of the attack and the allowlist, not a demonstrated capacity to resist. Second, the ordering does not transfer. Level 1, where the passport permits any recipient and the model's own judgment is the only thing in play, spans 0.0\% to 42.4\%; that ordering does not predict which models produced an unpermitted transfer at Levels 2 to 4. Third, the corpus is 4,371 attacks from an event with a \$6,500 prize pool, \$5,000 of it on Level 5 alone. It is not an upper bound on adversarial effort, and the concentration result of Section 5.1 shows how few sessions produced what we did observe.

The architectural claim is not that models are bad. It is that the deployer's residual risk after choosing the best model is an empirical unknown that varies by model, level and corpus, while the residual risk behind a correctly configured policy engine is bounded by the policy rather than by the model.

\subsection{What the zero is and is not}

The 0 of 69,297 is a property of a deterministic check evaluating a well-specified capability against an explicit allowlist. It is not a claim that authorization layers cannot fail. They fail when the policy is wrong, when the capability is specified too broadly, when the enforcement point can be bypassed, or when the action is not expressible as a policy decision. The result here says that when the policy is right and the check is in the execution path, a model that has been successfully manipulated into requesting an unpermitted transfer does not move money. That is a narrower statement than ``the system is secure'', and it is the statement the data supports.

It is also worth being explicit that the Level 5 zero is structural rather than empirical. A passport with no payment capability denies every payment by construction, so 0 of 20,109 there confirms the implementation works and demonstrates nothing about adversarial difficulty. The measured architectural effect is Levels 2 to 4.

\subsection{Implications for evaluation practice}

Three suggestions follow from the results rather than from principle. Report the enforcement architecture as an experimental variable, not as an implementation detail, because on this data it is the variable with the largest effect. Cluster on the source of attacks, because 47.9\% of our Levels 2 to 4 outcomes came from one session and any interval computed on evaluations would have been badly overconfident. And prefer domains with a checkable state change where possible: our judge panel disagreed with itself at kappa 0.167 at one level, and the only reason that is a footnote rather than a crisis is that nothing in the headline depends on it.

\subsection{Relation to the economics of agent autonomy}

Recent macroeconomic scenario work treats the automation share, the fraction of AI-performed task instances executed outright rather than reviewed by a person, as an exogenous technology parameter [15]. In deployment it is a per-action decision: does the agent execute, or does a person approve. Our data speaks to one input to that decision. Aggregate request rates are close in both architectures (Section 4.5), unpermitted transfers behind the layer are 0 of 69,297 within the bound of Section 4.3.1, and the policy denied 187 of the 25,640 calls it evaluated, so the great majority passed without a human in the loop (Section 4.6). We do not measure productivity, error costs outside this domain, or anything about non-payment actions, and we make no claim about the scenario parameters themselves.

\section{Limitations}

\textbf{Single domain, single tool.} Payments in a simulated bank with one transfer tool. Nothing here transfers to code execution, data access, or agent-to-agent delegation without new measurement.

\textbf{The attacker population is self-selected.} CTF participants competing for a \$6,500 prize pool are not general users and not state-grade adversaries. The rates are a best-effort expert adversarial result on one corpus, not a population estimate.

\textbf{Single run per cell.} Each cell is one pass at the CTF's production temperature (0.8 down to 0.1 by level), so per-cell rates are point estimates without a within-cell variance component. We chose fidelity to the conditions the attacks were written against over repeated sampling; a repeat-subset variance check was planned and is not in this release.

\textbf{Cross-provider decode defaults differ.} \texttt{top\_p} and penalty defaults are provider-specific and unnormalized. Model-to-model comparisons carry this confound.

\textbf{Level 1 and Level 5 are not attack-difficulty measurements.} Level 1 permits any recipient; Level 5 grants no capability and instructs the model to call the tool. Neither belongs in a headline and neither is reported as one.

\textbf{Partial multi-turn coverage.} Kimi K3 and GLM-5.3 have no behind-the-layer multi-turn cell and Qwen3.8 Max's is partial. Those models' behind-the-layer results rest on single-turn replay plus, where present, partial multi-turn data. The 0 of 69,297 is computed over completed cells only.

\textbf{The local engine is stricter than the published pack at Level 4.} It required an audit code in the memo that the public pack does not enforce (Section 3.10). The Level 4 zero is therefore evidence about the engine as configured here, not about the published pack's Level 4 configuration.

\textbf{The human validation slice was not completed.} It was pre-registered and the dataset card promised it here (Section 4.7). It is absent.

\textbf{Judge reliability is reported, not solved.} Kappa 0.167 at Level 3 and one judge's 64.4\% recall are disclosed rather than corrected. Any secondary analysis in the released data that uses the judge fields rather than the deterministic field inherits this.

\textbf{Attack-class coverage is uneven.} The corpus is dominated by authority claims and direct transfer requests. Tool-sequence exploits, parameter fuzzing and multi-agent delegation are underrepresented, because the CTF exposed a single-agent banking interface.

\textbf{Contamination is bounded, not excluded.} 36 prompts (86 attempts) match content that was publicly visible during the event, and a replay endpoint was served from April 11. Section 4.8 shows excluding the out-of-window tail changes nothing material, but we cannot rule out that some models' training data includes CTF discussion.

\textbf{No causal claim about model internals.} We observe that models request transfers at the same rate in both architectures. We do not claim to know why any individual model complied or refused.

\section{Reproducibility and release}

The evaluation set is released at \texttt{huggingface.co/datasets/aporthq/vault-benchmark-v1} under CC BY 4.0: 225,964 completed evaluation rows with per-row model, level, track, architecture, attack category, the executed tool calls, the deterministic outcome, the reconciled panel verdict, token usage and latency; the five level passports as enforced; the scoring code; the per-cell coverage and error counts; and the pre-registration with its two pre-analysis addenda.

Analysis in this paper is reproducible from the released rows with the script \texttt{release/paper\_analysis.py}, which reads the frozen snapshot \texttt{frozen-20260910T1230Z} when it is present and otherwise the released dataset. Bootstrap intervals use seed 20260915 and 1,500 resamples clustered on source session.

Five quantities cannot be recomputed from the released rows, and the script marks their tables unavailable rather than printing a number that would disagree with this paper: each judge's own pre-escalation verdict (so the inter-judge agreement in Section 3 needs the private snapshot), the counterfactual \texttt{would\_aport\_allow} field, turn-to-outcome, the deny reason codes, and banker cache tokens. Successful payments are 34 short at Levels 2 to 4 and 390 short at Level 5, every one of them GLM-5.3, whose transcripts are withheld under Z.ai's terms; the requests, recipient membership and unpermitted transfers those rows contribute are all present, so the headline figures reproduce exactly. Reading the released Parquet needs \texttt{pyarrow}; the frozen path needs only the standard library.

Withheld, and why: the attack transcripts are released in the evaluation rows, but the CTF harness, the judge prompts, and the signing keys are not, because releasing a working attack harness against a live policy configuration has a different risk profile from releasing the attacks themselves. The corpus carries a canary string; a reader who finds it in a model's output has found training-set contamination and we ask to be told.

Claims a reader can check from the public dataset alone: every count in Tables 1 through 7, the concentration and transfer results of Section 5, the judge agreement figures of Section 4.7, and the token and latency figures of Section 4.10. Claims that require our infrastructure: the parity of the local policy engine against the hosted verifier (Section 3.10) and the 53ms enforcement latency reported in [13].

\section{Conclusion}

We replayed 4,371 human-written attacks against 14 models in two architectures and recorded five separate events rather than one success rate. Adversarial replay elicited payment requests at rates spanning 0.1\% at Level 3 to 79.4\% at Level 4, differences between whole configurations rather than an isolated policy effect, and 99.4\% of the requests at Levels 2 to 4 recorded an allowlisted recipient. Behind a deterministic authorization check the models kept asking and kept being paid: 25,527 requests and 25,370 successful payments in 69,297 evaluations. What did not happen behind the check is the last row of the table. Unpermitted transfers were 140 of 76,842 with the model alone and 0 of 69,297 behind it, 105 against 0 on 68,970 matched triples, with a session-clustered upper bound of 0.38\%.

The check refused 187 of the 25,640 transfer calls it evaluated, 148 of them for a forbidden recipient, and allowed the other 25,453. That is the shape of the result: a boundary that enforces one specific restriction while the agent continues to work, not a mechanism that makes models refuse and not a general safety guarantee.

The structure behind the failures matters as much as their count. They came from two dozen sessions out of 790, one technique carried almost all of the multi-turn effect, and no attack defeated more than five of fourteen models. None of that is visible in a per-model success rate. What a deployer can take from it is narrow and, we think, useful: on this corpus the choice of model did not determine whether an action outside the policy executed, and a check in the execution path did. The Open Agent Passport specification that defines that check, and the enforcement latency it costs, are given in [13].

\section{Disclosure}

The author is the founder of APort Technologies Inc., which develops the authorization layer evaluated here. The benchmark was designed, run and analyzed by the author. This is a conflict of interest, and the mitigations are structural rather than declarative: the outcome metric is deterministic and computed from executed tool calls rather than from any judgment of ours; the pre-registration and its two pre-analysis addenda were filed before the run completed; the full evaluation set including every row that disagrees with our thesis is released; and the analysis script that regenerates every table in this paper from the frozen snapshot is released with it. A reader who distrusts the framing can recompute the numbers. We encourage independent replication, and we note that the strongest counter-argument to our conclusion (Section 6.1) is computed from our own data.

No human subjects data is released. CTF participants agreed to public release of attack content at registration; 7 submissions containing personal information were removed. No real money moved at any point.

\bibliographystyle{plain}

\begin{thebibliography}{18}
\bibitem{ref1}
E. Debenedetti, J. Zhang, M. Balunovi\'{c}, L. Beurer-Kellner, M. Fischer, and F. Tramèr. AgentDojo: A Dynamic Environment to Evaluate Prompt Injection Attacks and Defenses for LLM Agents. arXiv:2406.13352, 2024.
\bibitem{ref2}
Q. Zhan, Z. Liang, Z. Ying, and D. Kang. InjecAgent: Benchmarking Indirect Prompt Injections in Tool-Integrated Large Language Model Agents. arXiv:2403.02691, 2024.
\bibitem{ref3}
M. Andriushchenko, A. Souly, M. Dziemian, et al. AgentHarm: A Benchmark for Measuring Harmfulness of LLM Agents. arXiv:2410.09024, 2024.
\bibitem{ref4}
X. Zhang, Z. Wei, and H. Gong. MT-JailBench: A Modular Benchmark for Understanding Multi-Turn Jailbreak Attacks. arXiv:2605.11002, 2026.
\bibitem{ref5}
J. Song, X. Liu, and W. Yang. MultiBreak: A Scalable and Diverse Multi-turn Jailbreak Benchmark for Evaluating LLM Safety. arXiv:2605.01687, 2026.
\bibitem{ref6}
Y. Gao. How Reliable Is Your Jailbreak Judge? Calibration and Adversarial Robustness of Automated ASR Scoring. arXiv:2606.25487, 2026.
\bibitem{ref7}
S. Abdelnabi, A. Fay, A. Salem, et al. LLMail-Inject: A Dataset from a Realistic Adaptive Prompt Injection Challenge. arXiv:2506.09956, 2025.
\bibitem{ref8}
Y. Ruan, H. Dong, A. Wang, et al. Identifying the Risks of LM Agents with an LM-Emulated Sandbox. arXiv:2309.15817, 2023.
\bibitem{ref9}
D. Li, R. Zhang, et al. Preference Leakage: A Contamination Problem in LLM-as-a-Judge. arXiv:2502.01534, 2025.
\bibitem{ref10}
A. Panickssery, S. R. Bowman, and S. Feng. LLM Evaluators Recognize and Favor Their Own Generations. arXiv:2410.21819, 2024.
\bibitem{ref11}
A. Souly, Q. Lu, D. Bowen, et al. A StrongREJECT for Empty Jailbreaks. arXiv:2402.10260, 2024.
\bibitem{ref12}
P. Chao, E. Debenedetti, A. Robey, et al. JailbreakBench: An Open Robustness Benchmark for Jailbreaking Large Language Models. arXiv:2404.01318, 2024.
\bibitem{ref13}
U. Uchibeke. Before the Tool Call: Deterministic Pre-Action Authorization for Autonomous AI Agents. arXiv:2603.20953, 2026.
\bibitem{ref14}
K. Zhu, Y. Wang, and J. Wang. Open Payment Protocols for Autonomous Agents. arXiv:2602.20196, 2026.
\bibitem{ref15}
A. Korinek, C. Jones, S. Sacher, K. Cotter, and P. McCrory. Economic Scenarios for Transformative AI. Anthropic Institute Working Paper, 2026.
\bibitem{ref16}
S. Yao, N. Shinn, P. Razavi, and K. Narasimhan. $\tau$-bench: A Benchmark for Tool-Agent-User Interaction in Real-World Domains. arXiv:2406.12045, 2024.
\bibitem{ref17}
T. Hagendorff. Deception abilities emerged in large language models. Nature Communications 17:1435, 2026. DOI: 10.1038/s41467-026-69010-1.
\bibitem{ref18}
Z. C. Lipton and J. Steinhardt. Troubling Trends in Machine Learning Scholarship. arXiv:1807.03341, 2018.
\end{thebibliography}

\end{document}